\documentclass{article}
\usepackage{graphicx}

\begin{document}

\title{Stochastic equations and dynamics beyond mean-field theory}

\date{}

\author{Tommaso Rizzo \\ {\small ISC-CNR, UOS Rome, Universit\`a ``Sapienza'', } \\  {\small Piazzale A.~Moro 2, I-00185, Rome, Italy}
\\ {\small Dip.\ Fisica, Universit\`a ``Sapienza'',  }\\{\small Piazzale A.~Moro 2, I--00185, Rome, Italy } }

\maketitle

\begin{abstract}
The dynamical transition occurring in spin-glass models with one step of Replica-Symmetry-Breaking is a mean-field artifact that disappears in finite systems and/or in finite dimensions. The critical fluctuations that smooth the transition are described in the $\beta$ regime by dynamical stochastic equations. The quantitative parameters of the dynamical stochastic equations have been computed analytically on the 3-spin Bethe lattice Spin-Glass by means of the (static) cavity method and the equations have been solved numerically.
The resulting parameter-free dynamical predictions are shown here to be in excellent agreement with numerical simulation data for the correlation and its fluctuations.
\end{abstract}

The idea of a deep connection between structural glasses and spin-glasses (SG) with one step of Parisi's Replica symmetry breaking (1RSB) was put forward more than thirty years ago and has proven to be very influential \cite{kirkpatrick1987p,biroli2013perspective,wolynes2012structural}.
Mean-field SG models with 1RSB display a dynamical transition temperature $T_d$ where the Gibbs measure splits into an exponential number of equilibrium states,  {\it i.e. } there is a finite configurational complexity.  Often this is followed by a second (static) transition at $T_s$ where the configurational entropy vanishes.
The static transition  naturally evokes the Kauzmann temperature of supercooled liquids,  while the dynamical transition turns out to have the same qualitative features of the Mode-Coupling-Theory transition \cite{gotze2008complex}. 

While the existence of the Kauzmann temperature is controversial,  the MCT transition temperature is a very popular concept with both experimentalists and theorists,  indeed MCT captures many {\it qualitative} features of the physics of liquids upon supercooling,   notably two-step relaxation and stretched exponential decay.
Furthermore it  agrees {\it quantitatively} with numerical simulations \cite{nauroth1997quantitative,kob1999computer,sciortino2001debye,weysser2010structural}. 
Its main flaw is that in experiments one does not observe the sharp transition predicted by MCT but rather a  crossover from power-law to exponential increase of the relaxation time.
Many authors believe thus that it should be possible to fix MCT in some way although there is no agreement on how to do it. 

In the context of mean-field 1RSB SG one easily recognizes that the transition at $T_d$ is spurious due to their their mean-field nature and expects that ergodicity between $T_d$ and $T_s$ is restored in finite dimensions  by some activated processes: in practice one needs to go beyond mean-field dynamics.  On the other hand it seems that the nature of the problem is different for temperatures close to $T_d$ or deep in the MF glassy phase between $T_d$ and $T_s$.  In the following we will solely discuss progress made recently  for temperature close to $T_d$,  and we refer the reader to \cite{rizzo2021path} for recent work in the MF glassy phase. 

In order to go beyond MF and restore ergodicity one has to include fluctuations neglected at the MF level.  In 1RSB SG one sees that the dynamical transition has the features of a second-order phase transition \cite{franz2011field} and thus it is to be expected that the fluctuations are naturally described by a simple effective theory.  Due to certain non trivial features of the corresponding theory it turns out that it is equivalent to a set of dynamical stochastic equations called stochastic-$\beta$-Relaxation (SBR) equations in \cite{rizzo2014long,rizzo2016dynamical}.
In the following we will demonstrate the validity of SBR in 1RSB SG by comparing its predictions with numerical simulations for the paradigmatic Ising $p$-spin model.

We consider a system of $N$ spins each of which interacts with a fixed number $c=6$ of $p$-spin interactions with $p=3$ and evolve with Metropolis dynamics.  The (random) lattice is such that in the large $N$ limit loops are increasingly rare and it tends to the corresponding $c=6$ and $p=3$ Bethe lattice,  so that many thermodynamics quantities can be computed analytically by means of the cavity method \footnote{Note that instead of a random regular graph we generate the lattice by applying the $M$-layer construction \cite{altieri2017loop,rizzo2020solvable} to a triangular lattice with an interaction for each plaquette. This allows to obtain a tripartite graph enabling each set of spins to be updated at the same time.}.
The $p$-spin interactions are chosen randomly with values $J_{ijk}=\pm 1$ in the annealed ensemble.  Instead of the standard white average this corresponds  to weight each disorder instance with a factor proportional to the partition function of the model.  This is convenient for numerical studies because the averages over the interactions and the configurations can be exchanged,  in particular one can choose a random configuration and then generate the $J$'s accordingly  \cite{franz2011field,krzakala2011meltinga,krzakala2011meltingb}. 
The order parameter is the correlation with the initial condition:
\begin{equation}
C(t) = {1  \over N}\sum_i s_i(t)s_i(0) 
\end{equation}
In the thermodynamic limit the model displays a dynamical transition at a temperature $T_d$,  the correlation with the initial equilibrium configuration $C(t)$ approaches a plateau value $q_d$ with a power law
\begin{equation}
\langle C(t) \rangle \approx q_d+ \frac{1}{(t/t_0)^a}
\label{Betheg}
\end{equation} 
where the angle brackets mean average with respect to both the disorder and different thermal trajectories starting from the same initial configuration \cite{franz2011field}.
At finite $N$ one observes instead that even at $T=T_d$ the correlation deviates from the above mean-field expression and crosses the plateau value at a finite time that increases with the system size.
In order to describe this phenomenon we must compute corrections to mean-field theory.   Following the arguments and computations of \cite{rizzo2014long,rizzo2016dynamical} one can argue that close to $T_d$ the fluctuation of $g(t)$, defined as 
 \begin{equation}
g(t) \equiv C(t) - q_d\ ,
\end{equation}
 are described by SBR,  meaning that the generic $K$-point average obeys for $1\ll N < \infty$:
\begin{equation}
\langle g(t_1) \dots g(t_K) \rangle \approx [\hat{g}(t_1) \dots \hat{g}(t_K)]\, .
\label{equivalence}
\end{equation}
where  $\hat{g}(x,t)$ in the RHS is the solution of the SBR equations:
\begin{equation}
\sigma  + s =-\lambda \, \hat{g}^2(t)+{d \over dt}\int_0^t \hat{g}(t-s)\hat{g}(s)ds \ .
\label{SBRequa}
\end{equation}
The separation parameter $\sigma$ measures the distance from the critical point and vanishes at $T=T_d$. The square brackets mean average with respect to  the field $s(x)$ that is a quenched Gaussian random fluctuation of $\sigma$:  
\begin{equation}
[s]=0\, ,\ [s^2]=  \Delta \sigma^2  \ .
\end{equation}
the SBR equations have to be solved with the short-time condition
\begin{displaymath}
\lim_{t \rightarrow 0} \hat{g}(t) (t/t_0)^a=1
\end{displaymath}
where $\lambda$ and $a$ are related by the MCT relationship $\lambda={\Gamma^2(1-a) \over \Gamma(1-2a)}$.
In practice for times smaller than a Ginzburg time  $t_G \approx N^{1/(4a)}$  the observables on the LHS of eq. (\ref{equivalence}) can be accurately approximated with the values they have on the Bethe lattice while on times of order $t_G$ they are described by the RHS \cite{rizzo2020solvable}. This leads to the initial conditions of the SBR equations: the {\it short-time} behavior on times $O(t_G)$ matches the {\it long-time} behavior for times  $1 \ll t \ll t_G$, {\it i.e.} the mean-field result given by eq. {\ref{Betheg}}.

\begin{figure}[t]
\centering
\includegraphics[scale=1.2]{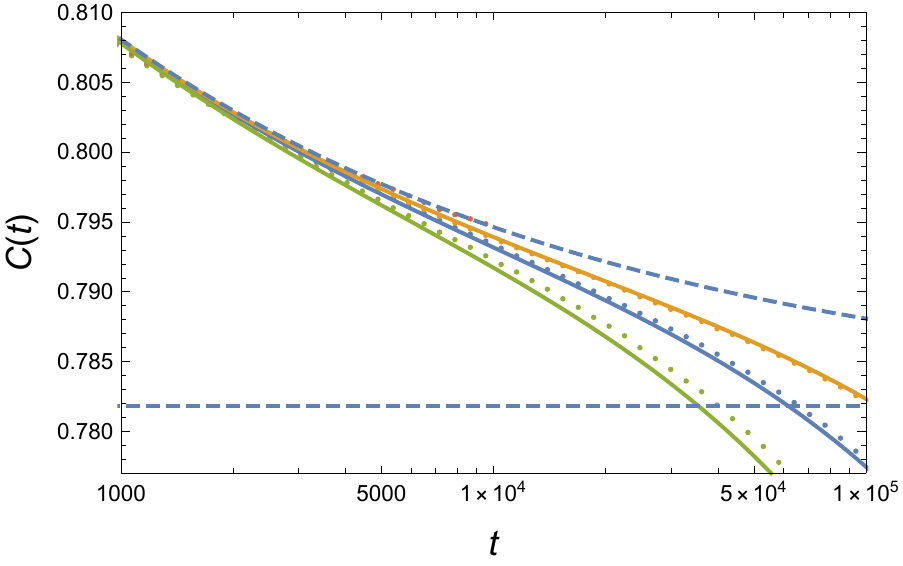} 
 \caption{Ising $p$-spin glass model on fixed connectivity lattice: average correlation with  initial equilibrium condition vs time at $T=T_d$.  Points from bottom to top: numerical data for $N=4.5 \times 10^5$,  $N=9 \times 10^5$,  $N=1.8 \times 10^6$ (Sample numbers are respectively 9554,  8048,  7701,  error bars are negligible on the scale of the plot).  The data follow the Bethe lattice  $N=\infty$ curve (dashed blue) at initial times  and deviate from it at later times increasing with $N$ eventually crossing the plateau value $q_d=0.78184$.  The solid lines are the corresponding SBR predictions describing the data when they start to deviate from the mean-field curve,  see text. }
\label{fig:plotC}
\end{figure}

\begin{figure}[t]
\centering
\includegraphics[scale=1.2]{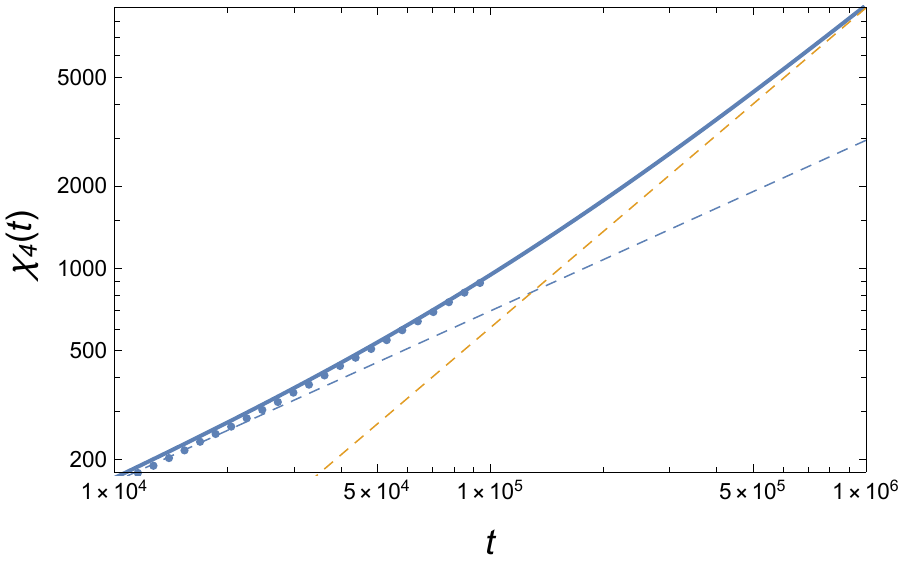} 
 \caption{Ising $p$-spin glass model on fixed connectivity lattice: $\chi_4(t)$ vs. time.  Points: numerical data for $N=1.8 \times 10^6$ (Sample number is 7701).  The data follow the mean-field asymptote $t^{2 a}$ (dashed blue) at initial times  and deviate from it at later times.  Solid: SBR prediction describing the data when they start to deviate from the mean-field short times asymptote $t^{2 a}$ (dashed blue).  SBR predicts a large times asymptotes  $t^{2b}$ (dashed,  yellow)}
\label{fig:plotX}
\end{figure}

Note that on the LHS of eq.  ( \ref{equivalence}) we have a model with a complex microscopic dynamics for which no analytic treatment of dynamics is available (not even in the fully connectec case),  on the RHS we have a (numerically) solvable set of equations  that were derived in \cite{rizzo2014long,rizzo2016dynamical} starting from symmetry considerations (essentially the detailed balance property of the dynamics) but {\it without} reference to any specific microscopic model.
The microscopic details however determine the actual values of the five SBR parameters $a$, $t_0$, $\Delta \sigma$ and $\sigma$ that  are needed to get quantitative predictions.
In order to obtain parameter-free predictions  these model-dependent parameters have been computed  analytically using existing \cite{franz2001exact, caltagirone2012critical,parisi2013critical,parisi2014diluted,lucibello2014one} and novel techniques based on the mean-field cavity method on the glassy phase of the transition (details elsewhere).
For the Bethe lattice with $3$-spin interactions $J=\pm 1$ and connectivity six we have thus obtained (both in the annealed and quenched ensemble) 
\begin{equation}
T_d=1.087815 \ , \ q_d=0.78184\ , \  a=0.31228\ , \ \Delta \sigma^2={0.02176 \over N } \\ . 
\label{barecouplings}
\end{equation}
The microscopic time-scale $t_0$ depends on the actual microscopic dynamics and therefore cannot be estimated by the static cavity method,  this is the only parameter that had to be extracted once and for all by fitting numerical data in the mean-field limit ({\it i.e. } at very large $N$) with the MF eq.  (\ref{Betheg}),   leading to $t_0=0.00866$.

Within SBR,  mean-field theory is recovered setting $\Delta \sigma^2=0$, in this case one recovers the critical MCT equation \cite{gotze2008complex},  in particular for $\sigma \geq 0$ ($T<T_d$) $C(t)$ never goes below the plateau value.  At finite $N$ there is instead a finite but small $\Delta \sigma$ so that the MCT transition is avoided and  $C(t)$ crosses the plateau at a finite time for all values of $\sigma$. 
In figure (\ref{fig:plotC}) we compare numerical data at $T=T_d$ ($\sigma=0$) with the SBR predictions that  were obtained solving numerically (by time discretization) eq. (\ref{SBRequa}) for many instances of the $s$'s. 
From the figure we note that the quality of the SBR predictions increases with $N$ and is excellent for $N=1.8 \times 10^6$,  especially considering that there is {\it no single fitting parameter} as even $t_0$ is estimated through an independent procedure.

SBR provides not only the average correlation but, according to eq. (\ref{equivalence}), also its  {\it fluctuations of all orders}.  To demonstrate this  in fig. (\ref{fig:plotX}) we compare data and theory for the $\chi_4(t)$ function that yields the fluctuations of the correlation:
\begin{equation}
\chi_4(t) \equiv N\,(\langle C^2(t)\rangle-\langle C(t)\rangle^2)\ .
\end{equation}
Within MF theory, at $T=T_d$  $\chi_4(t)$ should diverge with time as $t^{2 a}$ \cite{rizzo2020solvable}, instead on the Ginzburg time scale $t_G$ over which $C(t)$ deviates from MF and reaches the plateau value $q_d$,   $\chi_4(t) $ deviates from the MF  law and in the late $\beta$ regime follows a more pronounced $t^{2 b}$ growth where $b$ is related to $a$ by  $\Gamma^2(1+b) / \Gamma(1+2b)=\Gamma^2(1-a) / \Gamma(1-2a)$.

Our aim here was to demonstrate {\it quantitatively} that the theory is correct for 1RSB SG and we refer instead the reader to \cite{rizzo2015nature,rizzo2015qualitative} for a discussion of the rich phenomenology displayed {\it qualitatively} by SBR when considering the case of finite dimensions (in which $g(t)$ is promoted to a field $g(x,t)$) and the case of temperature slightly above and below $T_d$. 
Overall SBR predicts not only that the transition at $T_d$ is avoided as shown by fig. (\ref{fig:plotC}) or that the fluctuations deviate from mean-field theory as in fig. (\ref{fig:plotX}) but also that there is an essential qualitative change of in the structure of the fluctuations with the appearance of dynamical heterogeneities.
The excellent agreement between numerical data and the parameter-free predictions is reassuring because SBR is the natural theory for 1RSB SG, however it can be also obtained solely from the symmetries of the original dynamical problem:
this implies that the same description is potentially valid also for different models including notably supercooled liquids.  
 In \cite{rizzo2020solvable} its validity has been demonstrated in a class of Kinetically-Constrained-Models along the lines discussed here, by first computing analytically the coefficients of the theory and then comparing with numerical simulations.  Establishing the relevance of the theory and computing its parameters  for actual supercooled liquids  is a promising open problem.

\bibliographystyle{plain}   
\bibliography{rizzo}

\end{document}